\documentclass[aps,prx,twocolumn]{revtex4-2}

\usepackage[utf8]{inputenc}

\usepackage{xcolor}
\usepackage{amsmath}
\usepackage{amsfonts}
\usepackage{amssymb}
\usepackage{makeidx}
\usepackage{graphicx}
\usepackage{tikz}
\usepackage{float}
\usepackage{braket}

\usepackage[caption=false]{subfig}

\usepackage{color}
\usepackage[colorlinks=true,citecolor=blue,linkcolor=blue,urlcolor=blue]{hyperref}
\usepackage{url}

\newcommand{\be}{\begin{equation}}
\newcommand{\ee}{\end{equation}}
\newcommand{\bc}{\begin{center}}
\newcommand{\ec}{\end{center}}
\newcommand{\ben}{\begin{eqnarray}}
\newcommand{\een}{\end{eqnarray}}
\newcommand{\ave}[1]{\left< #1 \right>}

\newcommand{\op}[2]{| #1\rangle \langle#2|}
\newcommand{\la}{\langle}
\newcommand{\ra}{\rangle}

\newcommand{\Tr}{\text{Tr}}

\begin{document}

\title{An atom-doped photon engine:\\ Extracting mechanical work from a quantum system via radiation pressure}

\author{Álvaro Tejero}
\email{atejero@onsager.ugr.es}
\affiliation{Electromagnetism and Condensed Matter Department and Carlos I Institute for Theoretical and Computational Physics, University of Granada, E-18071 Granada, Spain.}
\author{Daniel Manzano}
\email{manzano@onsager.ugr.es}
\affiliation{Electromagnetism and Condensed Matter Department and Carlos I Institute for Theoretical and Computational Physics, University of Granada, E-18071 Granada, Spain.}
\author{Pablo I. Hurtado}
\email{phurtado@onsager.ugr.es}
\affiliation{Electromagnetism and Condensed Matter Department and Carlos I Institute for Theoretical and Computational Physics, University of Granada, E-18071 Granada, Spain.}

\date{\today}

\begin{abstract}
The possibility of efficiently converting heat into work at the microscale has triggered an intense research effort to understand quantum heat engines, driven by the hope of quantum superiority over classical counterparts. In this work, we introduce a model featuring an atom-doped optical quantum cavity propelling a classical piston through radiation pressure. The model, based on the Jaynes-Cummings Hamiltonian of quantum electrodynamics, demonstrates the generation of mechanical work through thermal energy injection. We establish the equivalence of the piston expansion work with Alicki's work definition, analytically for quasistatic transformations and numerically for finite time protocols. We further employ the model to construct quantum Otto and Carnot engines, comparing their performance in terms of energetics, work output, efficiency, and power under various conditions. This model thus provides a platform to extract useful work from an open quantum system to generate net motion, and it sheds light on the quantum concepts of work and heat.
\end{abstract}

\maketitle

\section{Introduction} Quantum heat engines, first proposed in \cite{scovil59a}, convert heat into work using a quantum system as a working medium. The growing interest in these quantum machines, fueled by the possibility of a true quantum advantage over their classical counterparts, has led to a fruitful field of research, intertwined with the current developments in quantum thermodynamics \cite{vinjanampathy:cp16}. Since the pioneering work of Alicki \cite{alicki:jpa79}, multiple models of heat engines operating in the quantum realm have been proposed, see, e.g., \cite{scully:science03,linden:prl10,scully:pnas11,rossnagel14a,hardal15a,uzdin:prx15,roulet:pre17,seah:njp18,zheng:pre16,binder_18,bhattacharjee:epjb21,rossnagel:science16,maslennikov19a,lindenfels19a,klatzow19a,ryan08a,peterson19a,ono20a,koski14a} and references therein. Although these models are different in nature, they are all governed by the laws of quantum mechanics, and their main objective is to efficiently convert heat into work in microscopic systems subject to both thermal and quantum fluctuations, and possibly in the presence of coherences. Advancements in technology have enabled pioneering experimental realizations of microscopic heat engines, operating either in the classical or quantum limits. Examples include machines made with single trapped ions \cite{rossnagel:science16,maslennikov19a,lindenfels19a}, nitrogen-vacancy centers in diamond \cite{klatzow19a}, spin systems \cite{ryan08a,peterson19a,ono20a}, or a single electron \cite{koski14a}, but also thermal engines built using colloidal particles \cite{martinez:sm17}. These studies have offered interesting results on the possibility of harnessing quantum effects to obtain advantages in these devices \cite{bhattacharjee:epjb21}. 

Concurrently with the development of quantum heat engines, there has been a theoretical discussion regarding the nature of heat and work in open quantum systems. The standard definition of heat and work, introduced by the seminal work of Alicki \cite{alicki:jpa79}, offers a primarily mathematical perspective (see also \cite{vinjanampathy:cp16}). This definition has been widely employed in the studies mentioned above on quantum thermal machines \cite{scully:science03,linden:prl10,scully:pnas11,rossnagel14a,hardal15a,uzdin:prx15,zheng:pre16,binder_18,bhattacharjee:epjb21,rossnagel:science16,maslennikov19a,lindenfels19a,klatzow19a,ryan08a,peterson19a,ono20a,koski14a}, but also to design novel devices such as quantum batteries \cite{alicki:pre13}, and to investigate fundamental issues such as, e.g., fluctuation theorems \cite{mukamel:prl03, campisi:rmp11, manzano:prb14, manzano:av18, micadei:prl20, manzano:njp21} and nonequilibrium thermodynamics \cite{manzano:pre12,asadian:pre13,manzano:njp16}. However, this consensus has also been challenged by work definitions based on ergotropy \cite{opatrny:prl21,seah:njp18} and population inversion \cite{brunner:pre12}. This suggests to study the  notions of work and heat in a quantum setting that allows the unambiguous identification of both magnitudes. In addition, and beyond these theoretical discussions, an important issue regarding quantum heat engines is how to actually \emph{extract} useful work from them to generate net motion. We address both challenges below.

In this paper, we present a model of an open quantum system that can be used to perform genuine mechanical work. The model is based on an atom-doped optical quantum cavity which drives a classical piston through radiation pressure. The atom-cavity system is described in terms of the paradigmatic Jaynes-Cummings model of quantum electrodynamics \cite{jaynes:procieee63,meher:epjp22}. We demonstrate that the injection of thermal energy into either the cavity or the atom, via an appropriate dissipative coupling to heat baths, can effectively generate work through piston displacements. We show analytically for quasi-static transformations that this (unambiguous) expansion work is equivalent to Alicki's work definition in the weak atom-radiation coupling limit, while numerical results suggest that the equivalence extends also to finite-time operations. We then use this setting to build quantum Otto and Carnot engines, analyzing in detail their performance. In particular, we obtain energy-frequency and pressure-volume cycle diagrams, and we compare the energetics, work output, efficiency, and power of both engines in the quasi-static limit and under finite-time protocols. Finally, the role of time-asymmetric stroke protocols and the accelerated motion of the piston are investigated using efficiency as a figure of merit. Note that, even though we focus mostly in work production, the model presented here can be modified to build up a refrigerator capable of cooling down a quantum system \cite{linden:prl10}. 

The paper is organized as follows. Section \ref{sec:model} presents the model and its open dynamics as described by a master equation. In Sec. \ref{sec:work}, we discuss the expansion work output of the pressure-driven piston, and compare it with Alicki's definition for quasi-static evolutions, proving their equivalence in the weak-coupling limit. Subsequently, Otto and Carnot's cycles are described in detail in Sec. \ref{sec:cycles}, and their numerical analysis both for quasi-static and finite-time evolutions is presented in Sec. \ref{sec:analysis}. Finally, we discuss our results in Sec. \ref{sec:conclusions}.

\begin{figure}[t]
\bc
\includegraphics[width=\linewidth]{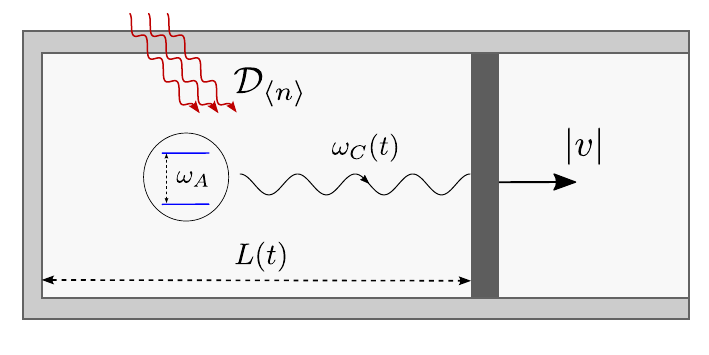}
\ec 
\vspace{-0.5cm}
\caption{An optical quantum cavity of length $L(t)$ and mode frequency $\omega_C(t)\sim L(t)^{-1}$, doped with a two-level atom of frequency $\omega_A$, drives a classical piston at speed $|v|$ through radiation pressure. We model the atom-cavity system using the paradigmatic Jaynes-Cummings model. Thermal energy is injected into the system via dissipative coupling $\mathcal{D}_{\ave{n}}$ to either the cavity or the atom.}
\label{fig:model}
\end{figure}

\section{Model and Master Equation}
\label{sec:model}

Our model is based on a two-level atom placed in an optical cavity of length $L$, as illustrated in Fig. \ref{fig:model}. The atom is coupled to an optical mode, which in turn interacts with a mobile classical wall through radiation pressure, resulting in a time-dependent cavity length $L(t)$. The interaction between the atom and the mode is described by the paradigmatic Jaynes-Cummings Hamiltonian, given by $H(t)=H_A+H_C(t)+H_I(t)$, where $H_A$, $H_C(t)$, and $H_I(t)$ represent the atom, cavity, and interaction Hamiltonians at time $t$, respectively \cite{jaynes:procieee63,meher:epjp22}. Note that the atom Hamiltonian is the only term that is time-independent, as it does not depend on the cavity length.

Adopting natural units hereafter, so that $\hbar=c=1$, the different Hamiltonian components are 
\ben
H_A &=& \frac{1}{2} \omega_A \sigma_z,\nonumber \\
H_C(t) &=& \omega_C(t) \left(a^\dagger a + \frac{1}{2}\right), \label{hamilt} \\
H_I(t) &=& \Omega(t) \left(a \sigma_+ + a^\dagger \sigma_-\right). \nonumber
\een
In these expressions, the operators $\sigma_z$ and $\sigma_\pm$ are the well-known Pauli operators acting on the atom, while $a$ and $a^\dagger$ correspond to the bosonic operators of the mode field. The parameter $\omega_A$ represents the natural frequency of the atom, while the resonant frequency $\omega_C(t)$ of the mode is inversely proportional to the cavity length \cite{garrison_08, gerry_04}, i.e. $\omega_C(t)= \alpha_0 / L(t)$, where $\alpha_0=\omega_C(0) L(0)$ is a constant defining the initial mode frequency. The coupling between the atom and the mode, denoted by $\Omega(t)$, is typically weak under experimentally-feasible settings \cite{garrison_08, gerry_04}, and satisfies $\Omega(t) \ll \omega_A, \omega_C(t)$, a condition that ensures the validity of the rotating wave approximation at all times \cite{breuer_02,garrison_08}, in agreement with experiments \cite{toyoda:prl13}.

Due to the dipolar character of the atom-radiation interaction, this coupling is also inversely proportional to the cavity size and takes the form $\Omega(t)=\kappa \omega_C(t) = \kappa\alpha_0 / L(t)$, where $\kappa \ll 1$ \cite{garrison_08, gerry_04}. To stress the experimental feasibility of this model, let us mention that possible experimental realizations of similar single-atom quantum engines have been recently proposed  \cite{barontini:njp19,hewitt:arxiv2023}, using e.g. an ultracold potassium (K) atom as the dopant/working medium, a thermal atomic cloud of rubidium (Rb) atoms as dissipative thermal baths, and optical tweezers as a piston model, though other atom combinations are also possible \cite{barontini:njp19}.

Note that the cavity Hamiltonian $H_C(t)$ in Eq.~\eqref{hamilt} does include a constant zero-point energy typical of the harmonic-oscillator Hamiltonian. This term is customarily neglected in the Jaynes-Cummings model as it amounts to a constant shift in the energy spectrum, with no direct effect on the dynamics. We prefer to keep it here, however, since this term will become relevant later on when comparing Alicki's work definition with the expansion work.

Our system is also driven by an external bosonic heat bath, which is modeled by a Lindblad term \cite{lindblad:cmp76}. In the Markovian limit, the dynamics of the atom-doped optical cavity is hence described by a master equation of the form \cite{breuer_02,manzano:aip20}
\be
\dot{\rho}(t) =  -\textrm{i} \left[ H(t),\rho(t) \right] + \mathcal{D}_{\ave{n}}\left[ \rho(t)\right],
\label{eq:me}
\ee
with a dissipative term 
\ben
 && \mathcal{D}_{\ave{n}}  \left[ \rho(t)\right] =\Gamma  \la n\ra \left(   A^{\phantom{\dagger}}_+ \rho(t) A_+^\dagger -\frac{1}{2} \left\{ \rho(t), A_+^\dagger A_+^{\phantom{\dagger}} \right\} \right) \nonumber \\
 && +  \Gamma  \left(\la n\ra+1\right) \left( A^{\phantom{\dagger}}_- \rho(t) A_-^\dagger -\frac{1}{2} \left\{ \rho(t), A_-^\dagger A_-^{\phantom{\dagger}} \right\} \right). 
\label{eq:diss}
\een
The constant $\Gamma$ represents the coupling strength between the system and the bath, and controls the relaxation time to the steady-state,
while $\la n\ra$ denotes the average number of resonant photons in the bath.
This temperature-dependent number dictates the occupation distribution of the different energy levels, independently of their spacing. Hereafter, we will characterize any thermal bath by a given value of $\la n\ra$ (and not by temperature), as the former parameter fully determines the steady-state behavior while being closer to experimental realizations of dissipative couplings. The generic operators $A_+$ and $A_-$ correspond to the jump operators associated with the system absorbing or emitting a photon from or to the bath, respectively, and their actual nature will depend on how the dissipative coupling is performed. In particular, to investigate the quantum conversion of thermal energy into mechanical work, we consider two scenarios. A first case consists in coupling the bath to the optical mode, implying $A_-=a$ and $A_+=a^\dagger$. This arrangement enables a direct transfer of thermal energy from the bath to the cavity radiation, which in turn drives the piston's motion. An alternative possibility consists in connecting the bath directly to the atom, such that $A_-=\sigma_-$ and $A_+=\sigma_+$. In this situation, the bath energy must be transferred from the atom to the cavity via the spontaneous emission of photons in order to produce work. We also note that, in the derivation of the master equation \eqref{eq:me}, the Lamb-shift term of the Hamiltonian has been neglected, as is customarily done for this type of systems where such correction has been proven small \cite{rivas:njp10}.

We would like to emphasize that the proposed baths are \emph{thermal} in the sense that they can prepare a thermal-like steady state \cite{zheng:pre16}. For instance, if a bath characterized by a fixed $\la n\ra$ is coupled just to a cavity mode of frequency $\omega$, the steady state of the system would be 
\be
\rho^{\text{th}} = \sum_{n=0}^{\infty} \dfrac{\ave{n}^n}{(1 + \ave{n})^{n+1} } \ket{n} \bra{n},
\label{eq:thermal1}
\ee
where $\la n\ra$ is given for a photon reservoir by a Bose-Einstein distribution $\la n\ra=\left[\exp(\omega/(k_B T))-1\right]^{-1}$ with null chemical potential at an effective temperature $T$ ($k_B$ is the Boltzmann constant). 
In this way fixing $\la n\ra$ dictates the occupation of the different levels in the steady state.

The interaction between the system and the piston occurs through the radiation pressure $p$ generated by the field mode. In natural units, the radiation pressure is nothing but the magnitude of the Poynting vector $\mathbf{S}$ representing the power flow of the electromagnetic field, $p = \left|\mathbf{S}\right|$. For plane waves, the radiation pressure can hence be expressed as a function of the electric field as $p=\left|\mathbf{E}\right|^2$ \cite{jackson:98}. For a single mode with frequency $\omega$ and under the dipole approximation, the quantized electric field can be written as \cite{garrison_08, gerry_04}
\be
\mathbf{E}(t) = \textrm{i} \left( \frac{ \omega}{2 V(t)} \right)^{\frac{1}{2}} \left( a \textrm{e}^{-\textrm{i}\omega t} - a^{\dagger} \textrm{e}^{\textrm{i}\omega t}\right) \mathbf{e},
\ee
where $\mathbf{e}$ is a unit vector in the direction of the field, and $V(t)$ is the volume of the cavity at time $t$. The radiation pressure operator is then
\be
\pi (t): = \frac{ \omega} {2 V(t)} \left( a^\dagger a + a a^\dagger -aa\textrm{e}^{-\textrm{i}2\omega t} -a^\dagger a^\dagger\textrm{e}^{\textrm{i}2\omega t}  \right)  ,
\label{pressure0}
\ee
so that the average pressure in a mixed state characterized by a density matrix $\rho(t)$ at time $t$ is just $p(t)=\la \pi(t) \ra = \Tr\left[ \pi(t) \rho(t) \right] $.

\section{Mechanical work and equivalence with Alicki's definition}
\label{sec:work}

In open quantum systems, the current standard definition of heat and work was given by Alicki \cite{alicki:jpa79}. For the general case of a system coupled to an environment via a dissipator and with a time-dependent Hamiltonian, the rate of energy change can be decomposed as 
\ben
\frac{d\left< E(t) \right>}{dt} &=& \frac{d}{dt} \Tr \left[ H(t)\rho(t) \right]  \\
&=& \Tr\left[  \dot{H}(t) \rho(t)  \right] + \Tr\left[ H(t) \dot{\rho}(t)  \right].\nonumber
\een
Work and heat are thus customarily defined as \cite{alicki:jpa79}
\ben
W_{\text{Al}} :&=& - \int  \Tr\left[  \dot{H}(t) \rho(t)  \right] dt , \label{eq:alicki} \\
Q_{\text{Al}} :&=& \int \Tr\left[ H(t) \dot{\rho}(t)  \right] dt ,\nonumber
\een
so that variations in energy due to changes in the system Hamiltonian are associated with work, while energy shifts due to state changes are  associated with heat \cite{ahmadi23a}. Note that we adopt a convention in which positive work is done by the system against the environment. We stress however that these definitions, though natural, remain somewhat controversial as e.g. they do not consider entropy variations or the role of possible internal energy fluxes, see \cite{ahmadi23a,tonner:pre05,weimer:epl08}.

\begin{figure*}
\includegraphics[scale=0.45]{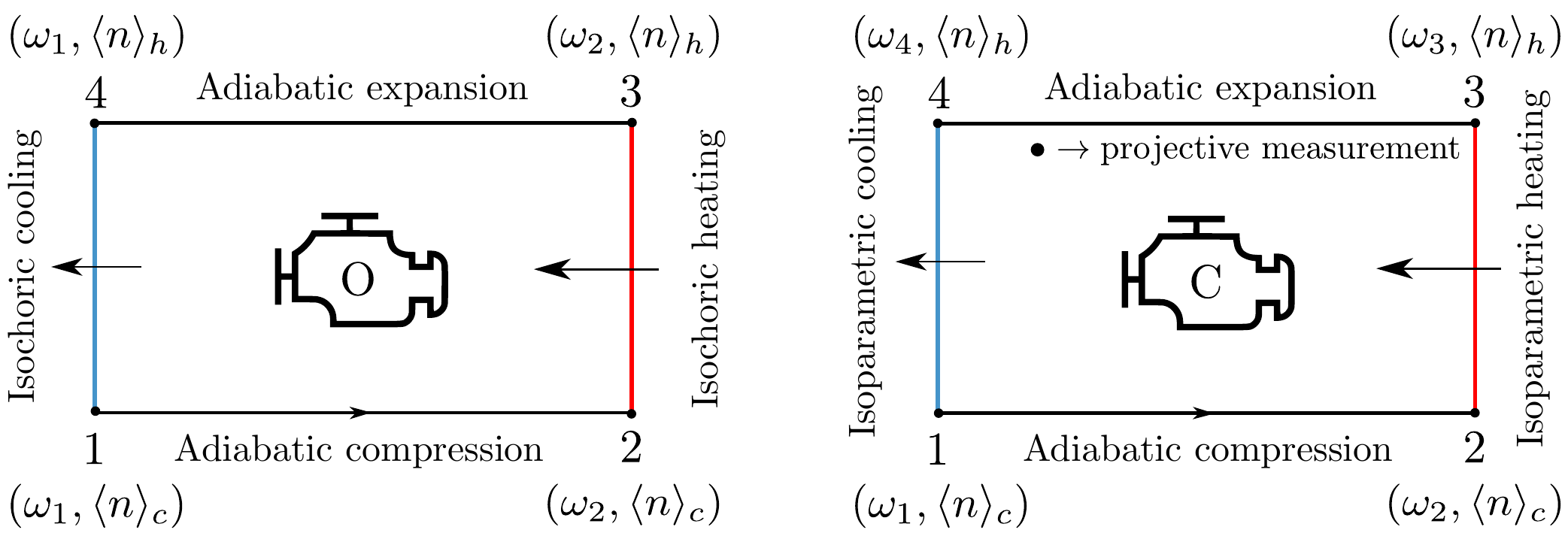}
\caption{(Color online) Sketch of the Otto (left) and Carnot (right) cycles, both of which consist in four strokes.
For the Otto cycle (left), the strokes are: (i) an adiabatic compression $L_1 \to L_2<L_1$ (equiv. $\omega_1 \to \omega_2>\omega_1$), (ii) an isochoric heating stroke in contact with a heat bath with a higher average photon number $\ave{n}_c \to \ave{n}_h>\ave{n}_c$, (iii) an adiabatic expansion stroke $L_2 \to L_1$ (or $\omega_2 \to \omega_1$), and (iv) an isochoric cooling stroke $\ave{n}_h \to \ave{n}_c$. For the Carnot cycle (right), the four strokes correspond to: (i) an adiabatic compression $L_1 \to L_2<L_1$, (ii) isoparametric (constant $\ave{n}$) expansion $L_2 \to L_3$ in contact with a hot bath with an average number of $\ave{n}_h$ resonant photons, (iii) an adiabatic expansion $L_3 \to L_4$, and (iv) isoparametric compression back to $L_1$ while in contact with a cold bath with $\ave{n}_c$ resonant photons on average. Projective measurements are performed after each stroke in all cases.}
\label{fig:ciclos}
\end{figure*}

A clear advantage of our optomechanical system is that work can be unambiguously identified with piston displacements, as in classical systems. This has also been recently implemented in some quantum rotor engine models \cite{roulet:pre17,seah:njp18}. Indeed, if the piston moves an infinitesimal length $\delta x$, the work performed can be written as $\delta W= F \delta x$, where $F$ is the generalized force acting on the piston. In our case this force is caused by the radiation pressure. Therefore, if the piston volume changes from $V_1$ to $V_2$ the average expansion work involved would be 
\be
W_{\text{exp}}=\int_{V_1}^{V_2}  \la p  \ra_t dV = \int_{V_1}^{V_2}   \Tr \left[\pi(t) \rho(t) \right]  dV.
\label{eq:expansion_work}
\ee
In order to compare this expansion work with Alicki's general definition, we first consider quasi-static transformations. In particular, if the above expansion is performed at a constant temperature $T$ and slowly enough to ensure that the system remains in a thermal equilibrium state at all times, i.e. $\rho(t)=\textrm{e}^{-\beta H(t)}/\mathcal{Z}(t)$ with $\beta=1/k_B T$ and $\mathcal{Z}(t)=\Tr[\textrm{e}^{-\beta H(t)}]$, we can calculate analytically the work output using Eq. (\ref{eq:expansion_work}).

For that, we first calculate the expected value of the pressure operator in the thermal state $\rho(t)$ using the spectral properties of the Jaynes-Cummings Hamiltonian $H(t)$. In particular, let $\ket{n,\pm}_t$ be the Hamiltonian eigenstates at time $t$, with associated eigenvalues $E_{n}^\pm (t)$. If $\ket{n}$ is the state of the cavity mode with $n$ excitations (Fock basis), the states $\ket{e}$ and $\ket{g}$ represent the excited and ground states of the atom, respectively. We consider the direct product basis of the whole atom+cavity system, so the Hamiltonian eigenbasis can be simply written as \cite{garrison_08, gerry_04}
\begin{align}
\begin{split}
\ket{n,+}_t =& \cos\left(\frac{\phi_n(t)}{2}\right)  \ket{n,e} + \sin\left(\frac{\phi_n(t)}{2}\right)  \ket{n+1,g}, \label{eigen} \\
\ket{n,-}_t =& -\sin \left(\frac{\phi_n(t)}{2}\right)  \ket{n,e} + \cos \left(\frac{\phi_n(t)}{2}\right)  \ket{n+1,g},
\end{split}
\end{align}
with $\phi_n(t)= \tan^{-1} \left( \frac{2 \Omega(t) \sqrt{n+1}}{\Delta (t)} \right)$, and $\Delta(t)=\omega_C(t)-\omega_A$ denotes the detuning between the cavity and the atom. Moreover, the Jaynes-Cummings Hamiltonian eigenvalues are
\be
E_{n}^\pm (t)= \left( n + \frac{1}{2} \right)  \omega_C(t) \pm  \left[\Delta(t)^2 + \kappa^2 \omega_C(t)^2 (n+1)\right]^{1/2} .
\ee
Using this eigenbasis together with the thermal form of $\rho(t)$ to compute the average pressure at time $t$, $\la p \ra_t =\Tr \left[ \pi(t)\rho(t) \right]$, see Eq. \eqref{pressure0}, and noting that the cavity volume can be written as $V(t)=S L(t)=S \alpha_0/\omega_C(t)$, with $S$ the piston's surface, we obtain 
\ben
\la p \ra_t &=& \frac{ \omega_C(t)^2}{2  \alpha_0 S  \mathcal{Z}(t)} \sum_{n=0}^\infty  \sum_{i=\pm} \bra{n,i}_t \textrm{e}^{-\beta H(t)} \left( a^\dagger a + a a^\dagger \right) \ket{n,i}_t  \nonumber \\
&=&  \frac{ \omega_C(t)^2}{ \alpha_0 S  \mathcal{Z}(t)} \sum_{n=0}^\infty \sum_{i=\pm} \left(n+\frac{1}{2}\right)  \textrm{e}^{-\beta E_n^{(i)}(t)} ,
\label{pressure}
\een
where we have already used that $\bra{n,\cdot}a a \ket{n,\cdot} = 0 =  \bra{n,\cdot}a^\dagger a^\dagger \ket{n,\cdot}$ for $\cdot=e,g$, see Eq.~\eqref{pressure0}. Note that there is a contribution of the vacuum zero-energy state to the average radiation pressure, which explains the reason why it is necessary to keep the zero-point energy term in the cavity Hamiltonian $H_C(t)$ of the Jaynes-Cummings model, see Eq.~\eqref{hamilt} above. In this way, writing now $dV=S \dot{L}(t) dt = -\alpha_0 S {\omega_C^{-2}}(t) \dot{\omega}_C(t)$, the expansion work reads
\ben
W_{\text{exp}} &=& \int_{t_1}^{t_2} \left< p \right>_t  S\frac{dL(t)}{dt}dt \\
&=& - \int_{t_1}^{t_2} dt  \frac{\dot{\omega}_C(t)}{ \mathcal{Z}(t)} \sum_{n=0}^\infty \sum_{i=\pm} \left(n+\frac{1}{2}\right)   \textrm{e}^{-\beta E_n^{(i)}(t)}  .\nonumber
\een
On the other hand, Alicki's work definition \eqref{eq:alicki} is based on the expected value of the time derivative of the Hamiltonian, $\dot{H}(t)= \dot{\omega}_C(t) \left[a^\dagger a + \frac{1}{2} + \kappa (a\sigma_+ + a^\dagger \sigma_-) \right]$. Using the previous spectral decomposition we obtain
\ben
&\Tr & \left[ \rho(t) \dot{H}(t)  \right] = \frac{ \dot{\omega}_C(t)}{\mathcal{Z}(t)}  \Big[ \sum_{n=0}^\infty \sum_{i=\pm} \left(n+\frac{1}{2}\right) \textrm{e}^{-\beta E_n^{(i)}(t)}  \nonumber \label{WAl2} \\
&+& \kappa \sum_{n=0}^\infty \sum_{i=\pm} \textrm{e}^{-\beta E_n^{(i)}(t)}\bra{n,i}_t (a\sigma_+ + a^\dagger \sigma_-) \ket{n,i}_t \Big]. \nonumber \\
 & &
\een
It is easy to show now,  using the particular form of the Hamiltonian eigenbasis, Eq.~\eqref{eigen}, that $\bra{n,\pm}_t (a\sigma_+ + a^\dagger \sigma_-) \ket{n,\pm}_t = \pm \sqrt{n+1} \sin(\phi_n(t)).$ Interestingly, in the weak coupling limit $\kappa\ll 1$ for the atom-cavity interaction in which the rotating-wave approximation leading to the Lindblad master equation~\eqref{eq:me} is valid, the angle $\phi_n = \tan^{-1} \left( \frac{2 \Omega(t) \sqrt{n+1}}{\Delta (t)} \right) \approx 2 \kappa \omega_C(t)\sqrt{n+1}/\Delta(t) + \mathcal{O}(\kappa^3)$, and therefore
\ben
\Tr & \left[ \rho(t) \dot{H}(t)  \right] & \underset{\kappa\ll 1}{=}  \frac{\dot{\omega}_C(t)}{\mathcal{Z}(t)}  \sum_{n=0}^\infty \sum_{i=\pm} \left(n+\frac{1}{2}\right)  \textrm{e}^{-\beta E_n^{(i)}(t)} \nonumber \\ 
&& + \mathcal{O}(\kappa^2)  .\
\een
This inmediately implies that Alicki's work can be written in this limit as
\ben
W_{\text{Al}}& \underset{\kappa\ll 1}  =  &- \int_{t_1}^{t_2}dt  \frac{\dot{\omega}_C(t)}{\mathcal{Z}(t)} \sum_{n=0}^\infty \sum_{i=\pm} \left(n+\frac{1}{2}\right)  \textrm{e}^{-\beta E_n^{(i)}(t)} \nonumber\\
&& + \mathcal{O}(\kappa^2)  .
\label{WAl1}
\een
We have hence proved that $W_{\text{Al}}=W_{\text{exp}} + \mathcal{O}(\kappa^2)$, so Alicki's definition and that of the expansion work are equivalent to first order in the weak-coupling limit $\kappa\ll 1$ in which the rotating-wave approximation is valid.

This equivalence thus ensures that our work definition is thermodynamically consistent \cite{vinjanampathy:cp16,alicki:jpa79,binder_18}. It is important to note, however, that the above proof of equivalence is limited by the approximations made, specifically by the assumption that the state of the system remains thermal throughout the entire process. In more complex scenarios, such as non-quasistatic evolutions, the two definitions might differ. However, as we shall see below, detailed numerical results comparing the expansion and Alicki's work outputs in finite-time operations strongly suggest that the equivalence of both work definitions extends also to these cases. Further analysis needs to be done to clarify the possible role of purely quantum effects, such as squeezing and coherences, where both definitions might not be equivalent. 

\section{Carnot and Otto Cycles}
\label{sec:cycles}

The model introduced in Sec. \ref{sec:model} and the equivalence demonstrated between the expansion and Alicki's work definitions allows us now to build finite-time quantum heat engines based on different types of thermodynamic cycles. Multiple models of quantum machines have been proposed in recent years \cite{linden:prl10,tonner:pre05,zheng:pre16,bhattacharjee:epjb21}, though in most cases an unsolved key issue has been how to actually extract useful work from them to generate motion. In this section we investigate the extraction of genuine mechanical work from the atom-doped quantum cavity optomechanical model introduced here. 

In particular, we will focus our attention on finite-time versions of the paradigmatic Carnot and Otto cycles, both of which consist of four different strokes, each of them of the same duration $\tau$ so the total cycle time is $t_\textrm{fin}=4\tau$. A schematic representation of both cycles can be seen in Fig.~\ref{fig:ciclos}, and note that the quasistatic case corresponds to the $\tau\to\infty$ limit. In particular, for the Otto cycle (i) the first stroke consists in an adiabatic (i.e., without heat exchange) compression which changes the system size from $L_1$ to $L_2<L_1$ in a time $\tau$, which corresponds to modifying the cavity mode frequency $\omega_C(t)\propto L(t)^{-1}$ from $\omega_1$ to $\omega_2>\omega_1$. This is followed by (ii) an isochoric (i.e., constant volume) heating stroke during which the system is in contact with a hot bath with $\la n \ra_h$ resonant photons on average, (iii) an adiabatic expansion stroke coming back to size $L_1$ (mode frequency $\omega_1$) in a time $\tau$, and (iv) an isochoric cooling stroke in contact with a cold reservoir with a photon parameter $\ave{n}_c<\ave{n}_h$ to close the cycle; see the left panel in Fig.~\ref{fig:ciclos}. On the other hand, for the Carnot cycle (i) the first stroke consists in adiabatic compression $L_1 \to L_2$ in a time $\tau$, followed by (ii) an \emph{isoparametric} (i.e., bath with constant $\ave{n}$) expansion $L_2 \to L_3$ of the same duration and speed during which the system is in contact with a hot bath with photon parameter $\ave{n}_h$. At this point, the bath is disconnected and (iii) for a time $\tau$ an adiabatic expansion $L_3 \to L_4$ continues at the same speed. Finally (iv), an isoparametric compression step is performed back to a cavity length $L_1$ while in contact with a cold reservoir with occupation number $\ave{n}_c$. 

Note that, in both cases (Otto and Carnot), the two adiabatic strokes involve unitary time-dependent evolutions, while the remaining two strokes have a dissipative nature as they involve energy exchange with thermal baths. Note also that after each stroke, a projective measurement is performed in the eigenbasis of the Hamiltonian to eliminate coherences, and ensure the repeatability of the cycle and that the energy output can be properly defined. As for the cycle initial state, we will start from a thermal product state for the cavity+atom system determined by an average occupation $\ave{n}_c$ for both subsystems, see below, and we will relax each engine to a steady cycle behavior by iterating the cycle many times before any measurement is performed. For quasistatic (i.e., large $\tau$) evolutions, the cycle's initial state will hence be a thermal state at the given temperature, but this thermalization is not guaranteed for finite-time protocols (or intermediate values of $\tau$), though we still observe convergence to a steady cycle for all cases.

To be more precise, the dynamics of the quantum system during a cycle can be described by the superoperator
\be
\mathcal{U}_{\text{cycle}}^{(\alpha)} = \prod_{k=1}^4 P_k \;\mathcal{U}_{k}^{(\alpha)},
\label{eq:strokes}
\ee
where $\alpha=\textrm{O(tto)},~\textrm{C(arnot)}$, and $P_k$ is the operator performing an energy projective measurement at the end of the stroke $k$; see below. In both type of cycles, the steps $1-2$ ($k=1$) and $3-4$ ($k=3$) correspond to adiabatic expansion/compression strokes. In these cases the evolution is unitary and given by 
\be
\mathcal{U}_{k}^{(\alpha)} \rho(t) = U_{k}^{\phantom{\dagger}}\rho(t) U_{k}^\dagger, 
\ee
with the evolution operator
\be
U_{k} = \exp\left( -\textrm{i} \int_{t_k}^{t_{k+1}} H(t) \; dt \right)  ,
\ee
and $t_k=(k-1) \tau$. To further advance, we need to reveal the time dependence of the cavity's length $L(t)$. A simple choice consists in assuming that the cavity wall moves at constant speed $|v|$ in the expansion ($v>0$) or compression ($v<0$) steps, so the cavity length changes linearly with time in between strokes, $L(t) = L_\textrm{ini} + vt$. Note that the quasistatic limit then corresponds to the $v\to 0$ case. This particular choice will of course influence our results on the system's efficiency, power output, and optimal parameters, and hence calls for further exploration of the space of possible expansion/compression protocols to gain deeper insights into the system's behavior, a task that we explore below. Note also that our choice for the expansion/compression (linear) protocol differs from previous works for heat engines based on ions in harmonic traps \cite{zheng:pre16}, where the mode frequency was chosen to vary linearly in time.

In addition to the unitary expansion and compression strokes, both the Carnot and Otto cycles include dissipative strokes characterized by energy exchanges with heat baths. These dissipative strokes differ, however, between the two types of engines. For the Otto cycle, these strokes are isochoric, meaning that the cavity volume (and thus the optical mode frequency) remain constant, i.e., $L(t)=L(t_k)$ and $\omega_C(t)=\omega_C(t_k)$ $\forall t\in[t_k,t_{k+1}]$ for the $2-3$ ($k=2$) and $4-1$ ($k=4$) Otto strokes. Consequently, the system dynamics during these Otto dissipative strokes can be described by the following superoperator
\be
\mathcal{U}_{k}^{(\textrm{O})} = \exp\left( \tau  \mathcal{L}_{k}^\textrm{(O)}\right),
\ee
with the following Lindbladian
\be
\mathcal{L}_{k}^\textrm{(O)} \rho(t) = -\textrm{i} [H(t_k),\rho(t)] + \mathcal{D}_{\ave{n}_{k}}\left[ \rho(t)\right]  ,
\label{LOtto}
\ee
where $\mathcal{D}_{\ave{n}_{k}}$ is the dissipator of the master equation (\ref{eq:diss}) with photon parameter $\ave{n}_k=\ave{n}_c,\ave{n}_h$ corresponding to the given cycle. 

In the Carnot cycle, the dissipative $2-3$ ($k=2$) and $4-1$ ($k=4$) strokes involve both energy exchanges with the baths and a simultaneous expansion/compression of the system. The superoperator capturing the evolution during these Carnot dissipative steps is thus
\be
\mathcal{U}_{k}^{(\textrm{C})} = \exp\left( \int_{t_k}^{t_{k+1}} \mathcal{L}_k^\textrm{(C)}(t) \; dt \right),
\ee
with a Carnot Lindblad superoperator of the form
\be
\mathcal{L}_k^\textrm{(C)}(t) \rho(t) = -\textrm{i} \left[ H(t),\rho(t) \right] + \mathcal{D}_{\ave{n}_{k}} \left[ \rho(t)\right] 
\ee
and a time-dependent Hamiltonian due to the evolution of the optical mode frequency $\omega_C(t)$.

Finally, in order to erase coherences, ensure the repeatability of the Otto or Carnot cycles, and allow for the determination of energy variations, a projective measurement in the energy eigenbasis is performed after each stroke, represented in Eq.~(\ref{eq:strokes}) by the operator $ P_k $. This measurement operator can be expanded as $ P_k =\sum_{n=0}^{\infty} \sum_{i=\pm} \op{n,i}{n,i}_{t_{k+1}}$. Consequently, if at a given time $t_{k+1}$ we have a density matrix $\rho(t_{k+1})$ and we perform the measurement, the post-measurement state will be given by
\be
\rho(t_{k+1}^+)=   P_k \rho(t_{k+1}) P_k  .
\label{eq:proj}
\ee
This measurement is required in order to define the energy variation and, therefore, the work production in each stroke. At the experimental level, this projection can be implemented in practice by measuring the level population using e.g. Raman sideband spectroscopy \cite{barontini:njp19,kaufman:prx12}. Specifically, in the Jaynes-Cummings model the energy levels can be measured using microwave spectroscopy \cite{lee:pra17}. This measurement process immediately removes any coherences from the Hamiltonian by making a direct projection into its eigenbasis.

\section{Numerical analysis}
\label{sec:analysis}

Once we have defined the details of the finite-time Otto and Carnot cycles, we are ready to analyze their work output and efficiency. To do so, we studied numerically, using the QuTiP Python library version 4.7 \cite{johansson:cpc13}, the time evolution of our quantum optomechanical system, as captured by the master Eqs. \eqref{eq:me} and \eqref{eq:diss} with the corresponding evolution superoperators \eqref{eq:strokes}-\eqref{eq:proj} for the different strokes. Note that, as discussed in Sec. \ref{sec:model}, the coupling to external heat baths in the dissipative strokes can be achieved by either (i) a direct transfer of thermal energy from the bath to the cavity radiation, or (ii) a coupling to the doping atom inside the cavity; see the definition of the dissipative term~\eqref{eq:diss} entering the master equation~\eqref{eq:me}. We will study both scenarios below.

\subsection{Initial state preparation}

The initial state of the system is chosen to be a direct product of thermal states for the cavity and the atom, both determined by the occupation number $\ave{n}_c$ of the cold bath. In particular, this state can be written as $\rho_{\textrm{ini}}= \rho_C(0) \otimes \rho_A = \frac{1}{\mathcal{Z}} \left( \textrm{e}^{-\beta_c H_C(0)} \otimes \textrm{e}^{-\beta_c H_A} \right)$, where $\beta_c$ would be  the inverse temperature of the bath and $\mathcal{Z}=\Tr \left[ \textrm{e}^{-\beta_c H_C(0)} \otimes \textrm{e}^{-\beta_c H_A} \right]$ is the partition function. The explicit expression of these thermal states is
\ben
\rho_C(0) &=& \sum_{n=0}^{\infty} \dfrac{\ave{n}_{c}^n}{(1 + \ave{n}_{c})^{n+1} } \ket{n} \bra{n}, \nonumber\\
\rho_A &=& \dfrac{1+\ave{n}_{c}}{1+2\ave{n}_{c}}  \ket{g}\bra{g} + \dfrac{\ave{n}_{c} }{1+2\ave{n}_{c}} \ket{e}\bra{e}, 
\label{eq:thermal}
\een
where we recall that $\ket{n}$ is the cavity state with $n$ excitations (Fock basis), $\ket{e}/\ket{g}$ are the excited/ground states of the atom, and $\ave{n}_{c}$ is the average number of resonant photons in the cold bath. It is important to notice that this initial state is not a thermal state for the interacting atom+cavity system, though it can be shown using fidelity measurements that it is indeed very close to such a thermal state. In any case, and in order to avoid dependence on initial state preparation, we let the different engines relax by iterating many times the cycle evolution before any measurement of energy and work output is performed. As mentioned above, for small (quasistatic) cavity wall velocities this protocol will lead to a thermal initial state for the cycle \cite{fan20a} associated with the given $\ave{n}_{c}$, but this thermalization is not guaranteed for finite-time processes (or higher speeds), though we still observe convergence to a steady cycle behavior in all cases studied. 


\begin{figure*}
\includegraphics[width=\linewidth]{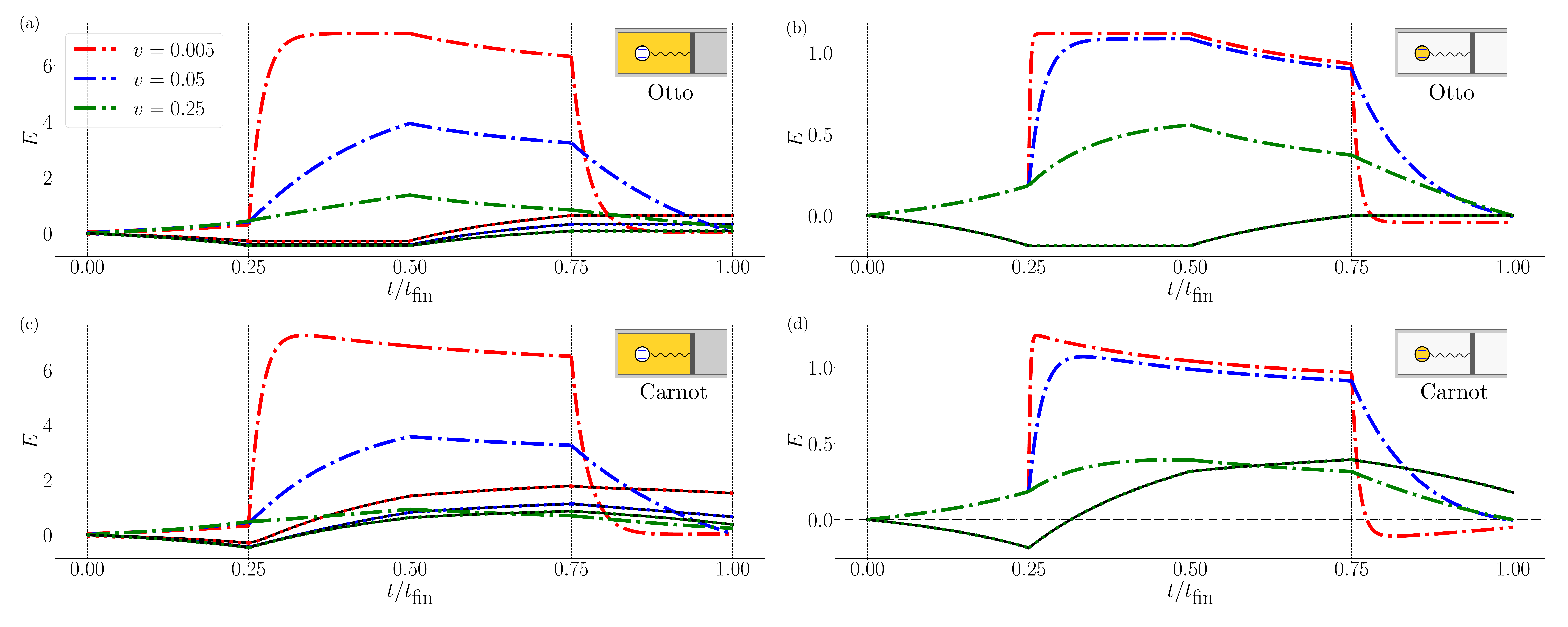}
\caption{(Color online) Energy variation and accumulated work output during a cycle of duration $t_\textrm{fin}=4\tau$ for different cavity wall speeds $|v|$.  (a) Otto engine with bath coupling to the cavity mode (colored cavity sketch);  (b) Otto engine with bath coupling to the atom (colored atom sketch); (c) Carnot engine with bath coupling to the cavity mode; and (d) Carnot engine with bath coupling to the atom. Dot-dashed lines correspond to energy variations, while dotted lines represent the expansion work output. Full black lines describe work measurements according to Alicki's definition, and note the agreement with the expansion work in all cases. Thin vertical lines represent the end of each stroke. In all panels the parameters are $\ave{n}_{c} = 5$, $\ave{n}_{h} = 20$, $\Gamma = 0.01$, $L_1=12.5$, $L_2=7.5$, $\omega_C(0) = \omega_A = 2\pi/L_1$, and $\kappa = 10^{-3}$. Note also that the legend in panel (a) applies to all panels.
} 
\label{fig:e_t}
\end{figure*}

\subsection{Energy balance and pressure-volume diagrams}

Figure~\ref{fig:e_t} shows the energy variation and the accumulated expansion work output as functions of time for both the Otto (top row) and Carnot (bottom row) engines, considering the two possible scenarios for the system-bath coupling, namely baths connected to either the cavity mode (left column, colored cavity sketch) or the atom (right column, colored atom sketch). For these particular examples, the cold bath is characterized by an average number of photons $\ave{n}_{c}=5$, while we choose $\ave{n}_{h}=20$ for the hot thermal bath. The system-bath coupling constant is set to $\Gamma=0.01$, see Eqs.~\eqref{eq:me} and \eqref{eq:diss}, and the constant $\kappa$ controlling the weak atom-radiation interaction strength $\Omega(t)$ is set to $\kappa=10^{-3}$, see Eq.~\eqref{hamilt}. In the Otto cycle the maximum cavity wall extension is chosen to be $L(0)=L_1=12.5$, while the minimum length is $L(\tau)=L_2=7.5$. Finally, the optical mode initial frequency is $\omega_C(0)=\alpha_0/L_1$, see the discussion below Eq.~\eqref{hamilt}, and we choose $\alpha_0=2\pi$ so that the extremal frequencies in the Otto cycle are $\omega_C(0)=\omega_1=2\pi/12.5\approx 0.503$ and $\omega_C(\tau)=\omega_2=2\pi/7.5\approx 0.838$. Moreover, the atom frequency $\omega_A$ is chosen to be the same as the initial optical mode frequency, $\omega_A=\omega_C(0)$. Note that Fig.~\ref{fig:e_t} includes results for different cavity wall speeds $|v|$, or equivalently different stroke times $\tau=(L_1-L_2)/|v|$, and the work sign convention is such that positive work corresponds to work produced by the moving cavity wall.

\begin{figure*}
\includegraphics[width=\linewidth]{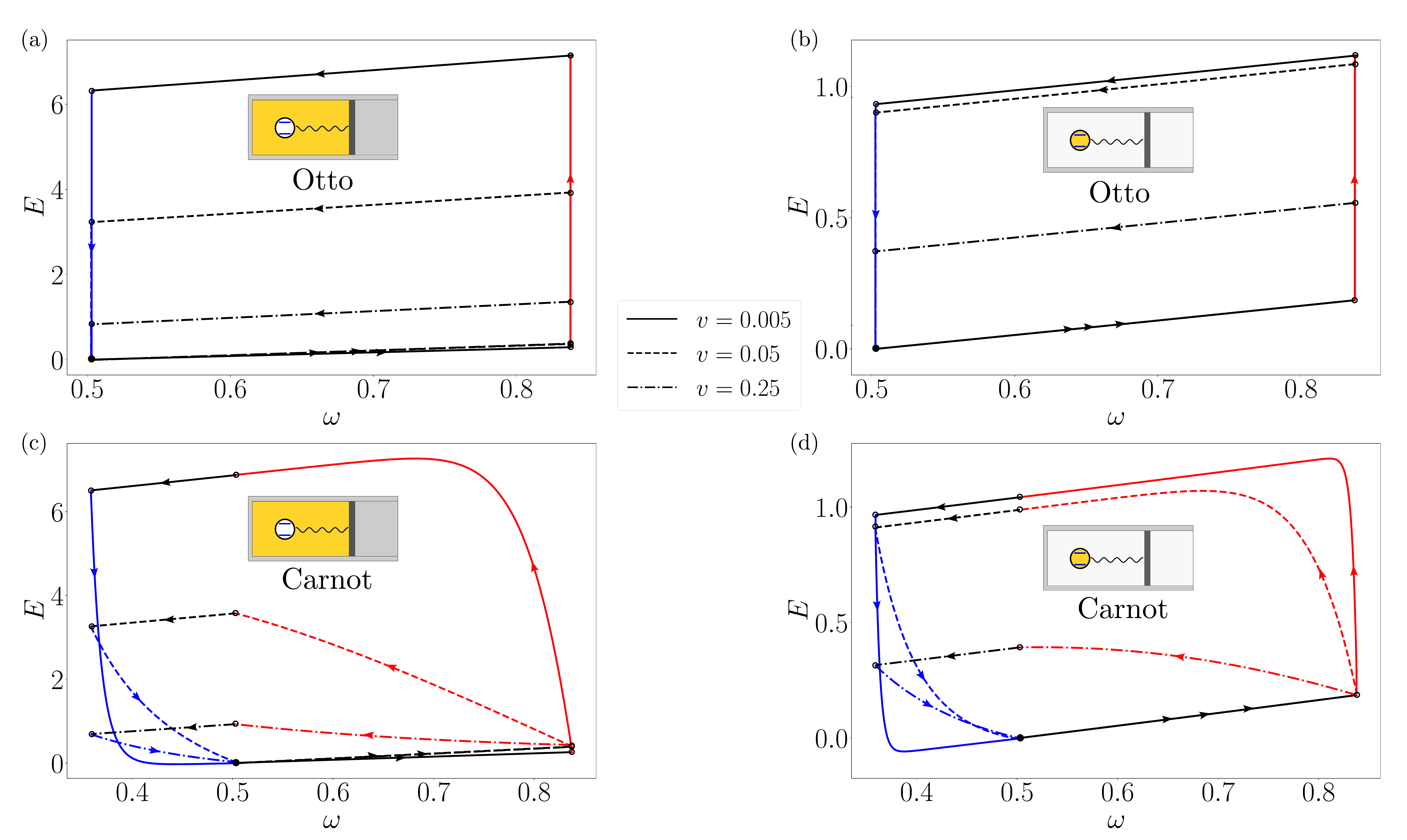}
\caption{(Color online) Energy as a function of the optical mode frequency $\omega$ during a cycle for different cavity wall speeds. (a) Otto engine with bath coupling to the cavity mode;  (b) Otto engine with bath coupling to the atom; (c) Carnot engine with bath coupling to the cavity mode; and (d) Carnot engine with bath coupling to the atom. Black lines represent adiabatic compression/expansion strokes, while the red (blue) line denotes the isochoric heating (cooling) stroke for the Otto case (top row, a-b), and the the isoparametric expansion (compression) stroke for the Carnot engine (bottom row, c-d). In all cases, the cycle initial state is marked with a full circle, while the initial point of all subsequent strokes is denoted with an open circle, and arrows signal the cycle direction. In all cases the parameters are $\ave{n}_{c} = 5$, $\ave{n}_{h} = 20$, $\Gamma = 0.01$, $L_1=12.5$, $L_2=7.5$, $\omega_C(0) = \omega_A = 2\pi/L_1$, and $\kappa = 10^{-3}$. Note also that the legend in panel (a) applies to all panels.
} 
\label{fig:e_omega}
\end{figure*}

We start with the Otto engine, whose dynamics is somewhat simpler to interpret. Indeed, in this case work is only involved during the adiabatic expansion and compression strokes, while all energy exchanges during the thermalization strokes are linked to heat transfer, in accordance with Alicki's definition of heat, see Eq.~\eqref{eq:alicki}. As shown in Figs.~\ref{fig:e_t}(a) and ~\ref{fig:e_t}(b), during the first Otto stroke (adiabatic compression) there is always a negative work production and a subsequent energy increase in the atom+cavity system. Note that the results of this first adiabatic compression stroke are mostly independent of the cavity wall velocity $v$ for both bath couplings. This is related to both the ideal character of the radiation filling up the cavity and the weak atom-radiation interaction, which causes both the energy and work evolution to depend exclusively on the initial and final cavity states, and not on how fast we push the cavity wall to move from one to the other. This is also apparent in the energy-frequency diagrams for the Otto engines, see Figs.~\ref{fig:e_omega}(a) and ~\ref{fig:e_omega}(b), where compression and expansion strokes are depicted by black lines (red and blue lines represent the heating and cooling strokes, respectively), and note that since $\omega_C(t)\propto L(t)^{-1}$, the compression stroke [$L(t)\downarrow$] corresponds to an increasing mode frequency  [$\omega_C(t)\uparrow$].

The next stroke is the isochoric heating of the system. This is achieved by letting either the cavity [Fig.~\ref{fig:e_t}(a)] or the atom [Fig.~\ref{fig:e_t}(b)] contact the hot heat bath with $\ave{n}_h$ resonant photons during a time $\tau$ at constant volume. Interestingly, the coupling to the hot thermal bath leads to a steep, rapid energy increase in both cases. For large stroke durations $\tau$ (equiv. small $|v|$), the working medium (atom+cavity system) has more time to drain energy from the hot bath, leading to a higher energy quasi-thermal state. Note however that a clear difference in the energy level reached appears at this point between the bath-cavity and bath-atom couplings for the same stroke duration. This is reasonable since the cavity radiation has an unbounded energy spectrum and hence a very large thermal capacity, while the atom has only two energy levels and thus a comparatively small thermal capacity. This particular feature also makes thermalization much faster when the baths are coupled to the atom in the Otto case, compare energy curves in Figs.~\ref{fig:e_t}(a) and ~\ref{fig:e_t}(b). The differences in total energy after the isochoric heating stroke are also apparent in the energy-frequency diagrams of Figs.~\ref{fig:e_omega}(a) and ~\ref{fig:e_omega}(b) (red lines). On the other hand, due to the isochoric character of this second stroke, no expansion work is involved in this step so that the accumulated work remains constant and negative. 

This unfavorable work balance, however, is compensated during the third Otto stroke (adiabatic expansion): the work invested initially in adiabatically compressing the cavity is surpassed by the work obtained from the adiabatic expansion from a higher energy state, see work (and energy) curves in Figs.~\ref{fig:e_t}(a) and ~\ref{fig:e_t}(b), leading to a positive accumulated work output at the end of the Otto cycle. During the adiabatic expansion, the system energy decreases from different energy states determined by the duration of the isochoric heating stroke, but the amount of energy invested in the expansion depends only very weakly on the cavity wall velocity, leading to small work differences for the varying velocities. This is specially acute for the bath-atom coupling, where work curves exhibit mostly unobservable variations with the cavity wall velocity. Finally, an isochoric cooling stroke follows to end the cycle by coming back to the original state. Again, the work accumulated during a cycle when the baths are coupled to the cavity radiation is much higher than the accumulated work for bath-atom coupling, due to the difference in thermal capacities between the cavity radiation and the two-level atom. Note also that the work output in the Otto cycle with bath-atom coupling is positive but very small as compared with the energy curves; see Fig.~\ref{fig:e_t}(b).

The case of the Carnot engine is slightly more involved, as there is work production in all stages of the cycle; see Figs~\ref{fig:e_t}(c) and ~\ref{fig:e_t}(d). In this case, the initial work invested in adiabatically compressing the cavity during the first stroke is compensated in the subsequent second and third Carnot strokes, corresponding to the isoparametric and adiabatic expansion steps, respectively. The energy curves for the Carnot case also display a steep, rapid increase/decrease when first put in contact the hot/cold thermal baths. However, and in contrast with the Otto case, these curves exhibit a remarkable overshoot during the isoparametric compression stroke and a corresponding undershoot during the isoparametric expansion stroke, the effect being most apparent (if present) for the slowest cavity wall velocity. Note also that these overshoots and undershoots lead to difficult relaxation for slow cavity wall velocities. The presence of these energy over/undershoots for some (small and intermediate) cavity wall velocities is also apparent in the energy-frequency diagrams for the Carnot cycle; see Figs.~\ref{fig:e_omega}(c) and ~\ref{fig:e_omega}(d). In particular the energy maxima and minima for $|v|=0.005,~0.05$ are reached for nontrivial frequency values inversely proportional to the size of the cavity at the time of the over/undershoot observed in the energy-time plots of Figs.~\ref{fig:e_t}(c) and ~\ref{fig:e_t}(d).

In Sec. \ref{sec:work} we demonstrated the equivalence of the expansion and Alicki's work definitions to first order in the atom-radiation interaction coupling, in the case of quasistatic transformations. The question remained as to whether this equivalence still holds for finite-time processes. To test this possibility, we measured the time dependence of the work output in the previous examples using also Alicki's definition based on the variations in energy due to changes in the system Hamiltonian; see Eq.~\eqref{eq:alicki}. These measurements are represented by thin black lines in Figs.~\ref{fig:e_t}(a)-~\ref{fig:e_t}(d), and note that in all cases Alicki's work definition fully agrees with the expansion work measurements. This shows that indeed the equivalence between the expansion and Alicki's work is robust under finite-time engine operations, both for the Otto and Carnot cycles.

In general, a comparison of both types of engines shows that, for a given cavity wall speed, the Carnot cycle yields a higher work output than the related Otto cycle. Indeed, the Carnot case has more freedom to optimize the work output, hence its higher performance. It is also interesting to confirm that the direct coupling of external baths to the doping atom leads to a net expansion work via the emission/absorption of energy quanta to/from the optical cavity radiation, which in turn increases radiation pressure, driving the cavity wall. This is an explicit example of conversion of quantum energy into mechanical work mediated by quantized radiation.

\begin{figure}
\includegraphics[width=\linewidth]{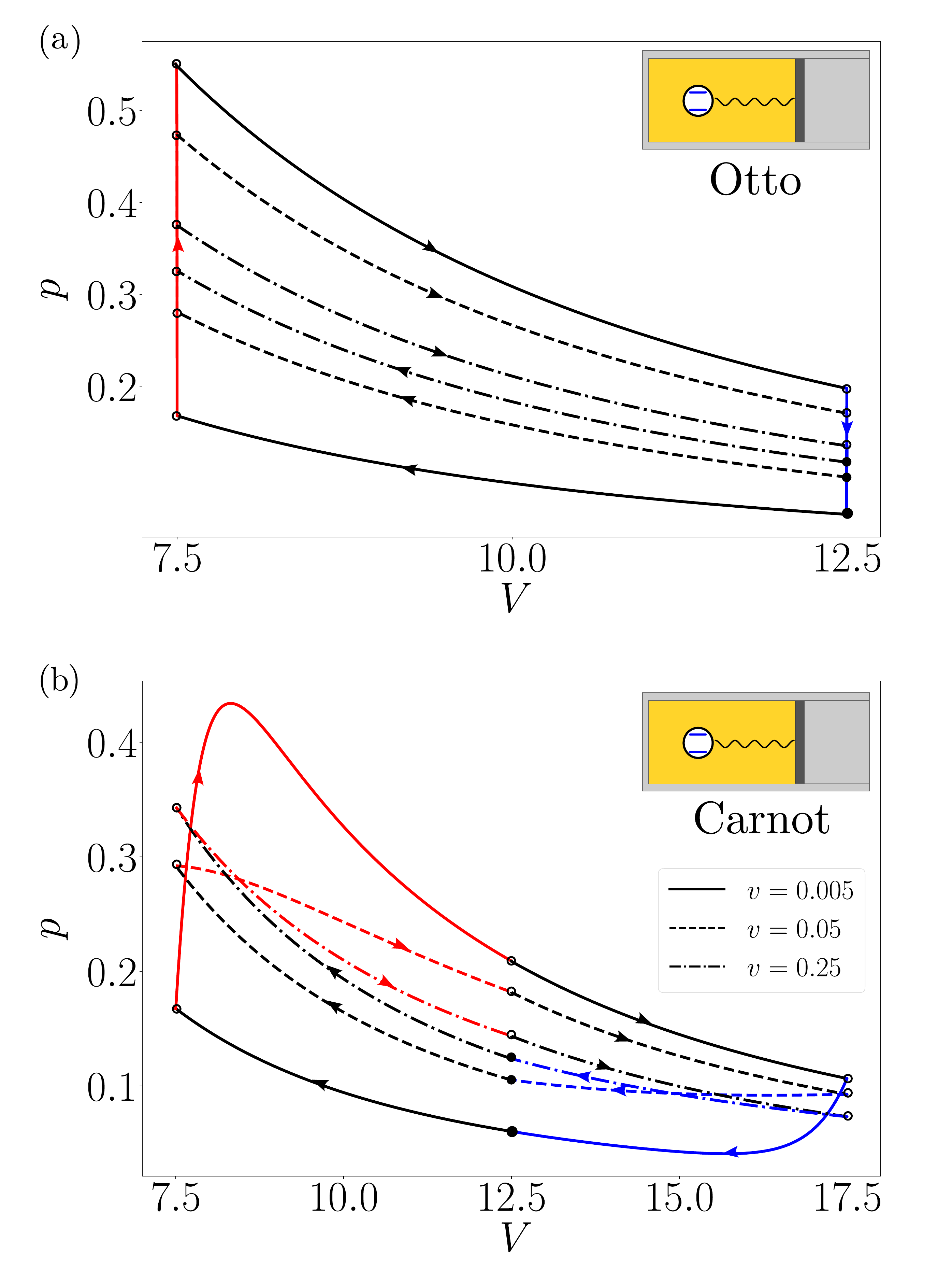}
\caption{(Color online) Pressure-volume diagrams for the Otto (top) and Carnot (bottom) cycles for a system with the thermal bath connected to the cavity mode and different cavity wall speeds. Black lines represent adiabatic compression/expansion strokes, while the red (blue) line denotes the isochoric heating (cooling) stroke for the Otto case (top panel), and the isoparametric expansion (compression) stroke for the Carnot engine (bottom panel). In all cases, the cycle initial state is marked with a full circle, while the initial point of all subsequent strokes is denoted with an open circle. The parameters are $\ave{n}_{c} = 5$, $\ave{n}_{h} = 20$, $\Gamma = 0.01$, $L_1=12.5$, $L_2=7.5$, $\omega_C(0) = \omega_A = 2\pi/L_1$, and $\kappa = 10^{-3}$. Note also that the legend in panel (b) applies to both panels.
} 
\label{fig:p-v}
\end{figure}

The design and versatility of our atom-doped engines, based on the radiation-induced displacements of the optical cavity wall, allows us to study also Clapeyron-style pressure-volume diagrams; see Fig.~\ref{fig:p-v}. In particular, this figure shows results for the thermal bath-cavity coupling case, though similar results (not shown) can be obtained for the bath-atom coupling scenario. Note that, as in the energy-frequency plots of Fig~\ref{fig:e_omega}, the area enclosed by the different pressure-volume cycles decreases as the velocity of the cavity wall increases, a clear indication of the decreasing work output as $|v|$ increases. Note also that, for the Carnot cycle and small cavity wall speed, there exist a pressure overshoot and undershoot at the volume points associated to the optical mode frequency where the system energy exhibits equivalent extrema; see Fig.~\ref{fig:e_omega}(c).

\begin{figure}
\centering
\includegraphics[width=8.7cm]{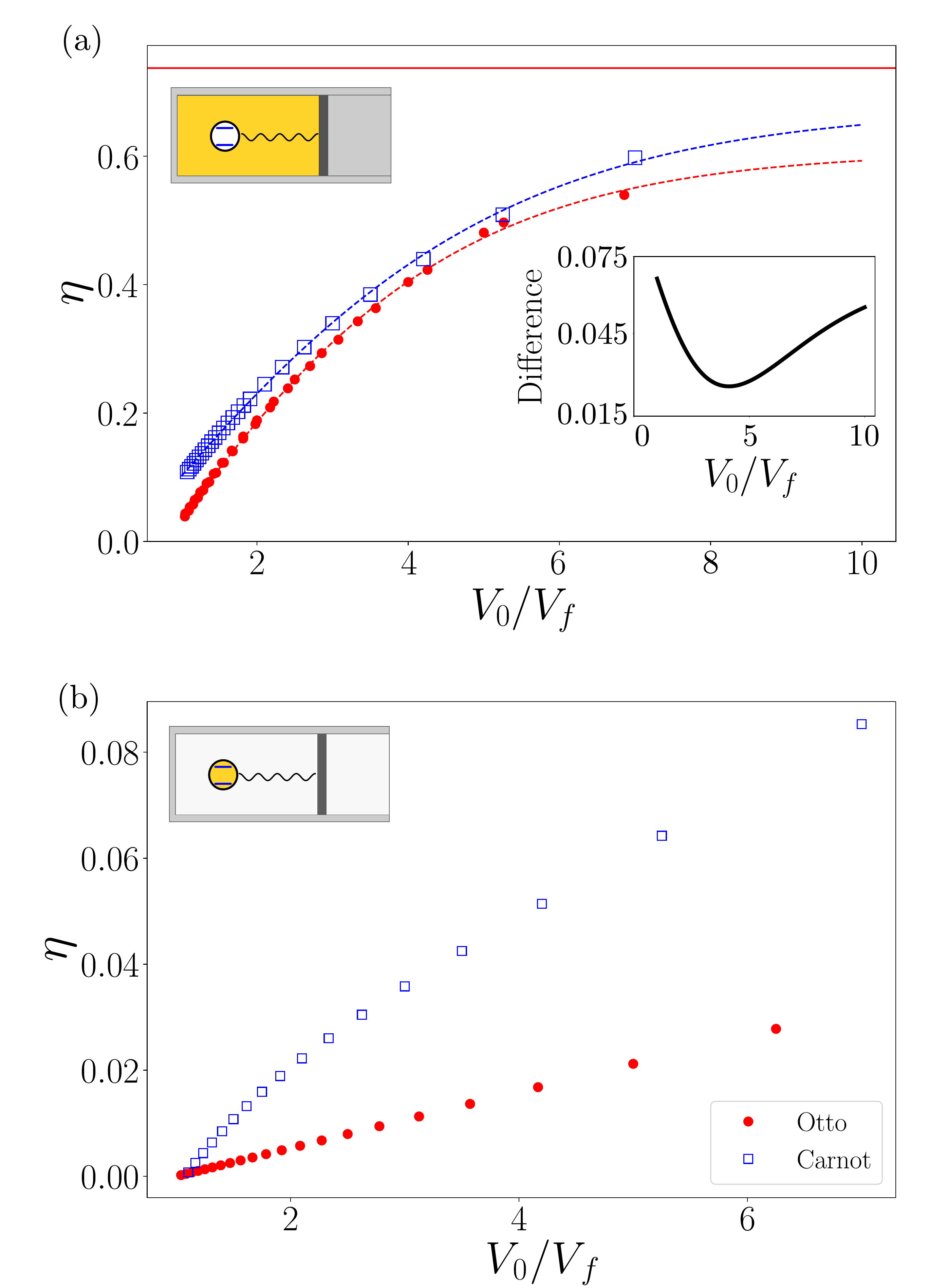}
\vspace{-0.3cm}
\caption{(Color online) Efficiency of Carnot ($\square$) and Otto ($\bullet$) cycles as a function of the compression ratio $V_0/V_f$. Top panel: Thermal baths coupled with the cavity. The red vertical line represents the Carnot theoretical limit, while the dashed lines are polynomial fits to the data. The inset shows the difference in efficiency between the Carnot (C) and Otto (O) cases, $(\eta_\textrm{C}-\eta_\textrm{O})$, as a function of $V_0/V_f$, which exhibits a clear minimum. Bottom panel: Thermal baths coupled with the atom. In all cases the parameters are $\ave{n}_{c} = 5$, $\ave{n}_{h} = 20$, $\Gamma = 0.01$, $L_1=12.5$, $\omega_C(0) = \omega_A = 2\pi/L_1$, and $\kappa = 10^{-3}$. Note also that the legend in panel (b) applies to both panels.} 
\label{fig:efficiency}
\end{figure}

\subsection{Engine efficiency and power output}

To properly compare both types of engine cycles and assess their performance, it is most reasonable to monitor the engine efficiency as a figure of merit. The efficiency of a motor is defined as $\eta=\frac{|W|}{|Q_H|}$, where $W$ represents the net work performed during a cycle and $Q_H$ denotes the heat transferred from the hot reservoir to the system (determined as the energy variation during this process that is not due to work). As in classical engines, we expect the efficiency to depend strongly on the compression ratio, i.e., the ratio between the maximum ($V_0$) and minimum ($V_f$) volumes reached by the cavity during the cycle. Figure ~\ref{fig:efficiency} shows the engine efficiency as a function of $V_0/V_f$ measured in the quasistatic (small $|v|$) limit for both the Carnot and Otto cycles, considering the thermal bath coupled to the cavity (top panel) or to the atom (bottom panel). A first observation is that for engines with direct heat exchange between the reservoir and the cavity mode (top panel in Fig.~\ref{fig:efficiency}), the efficiency is quite high and approaches steadily the Carnot limit (upper red line) for large compression ratios. On the other hand, for the bath-atom coupling case (bottom panel), the efficiency is an order of magnitude lower, though it maintains its monotonously increasing behavior with the compression ratio for both types of engines. This difference in magnitude is of course related to the limited heat capacity of the two-level atom as compared to the large thermal capacity of the cavity radiation. It is also interesting to note that, in all cases, the Carnot cycle exhibits a higher efficiency than the equivalent Otto cycle, though the relative difference is smaller when the thermal connection is made to the cavity mode. Moreover, while the (positive) difference in efficiency between the Carnot and Otto cycles increases monotonously as a function of $V_0/V_f$ for the bath-atom coupling, this difference exhibits a minimum at a non-trivial compression ratio in the case of bath-cavity thermal coupling; see the inset in the top panel of Fig.~\ref{fig:efficiency}. This defines a sort of optimal compression ratio for which the Otto efficiency approaches most closely the Carnot one for bath-cavity coupling.

\begin{figure*}
\includegraphics[width=\linewidth]{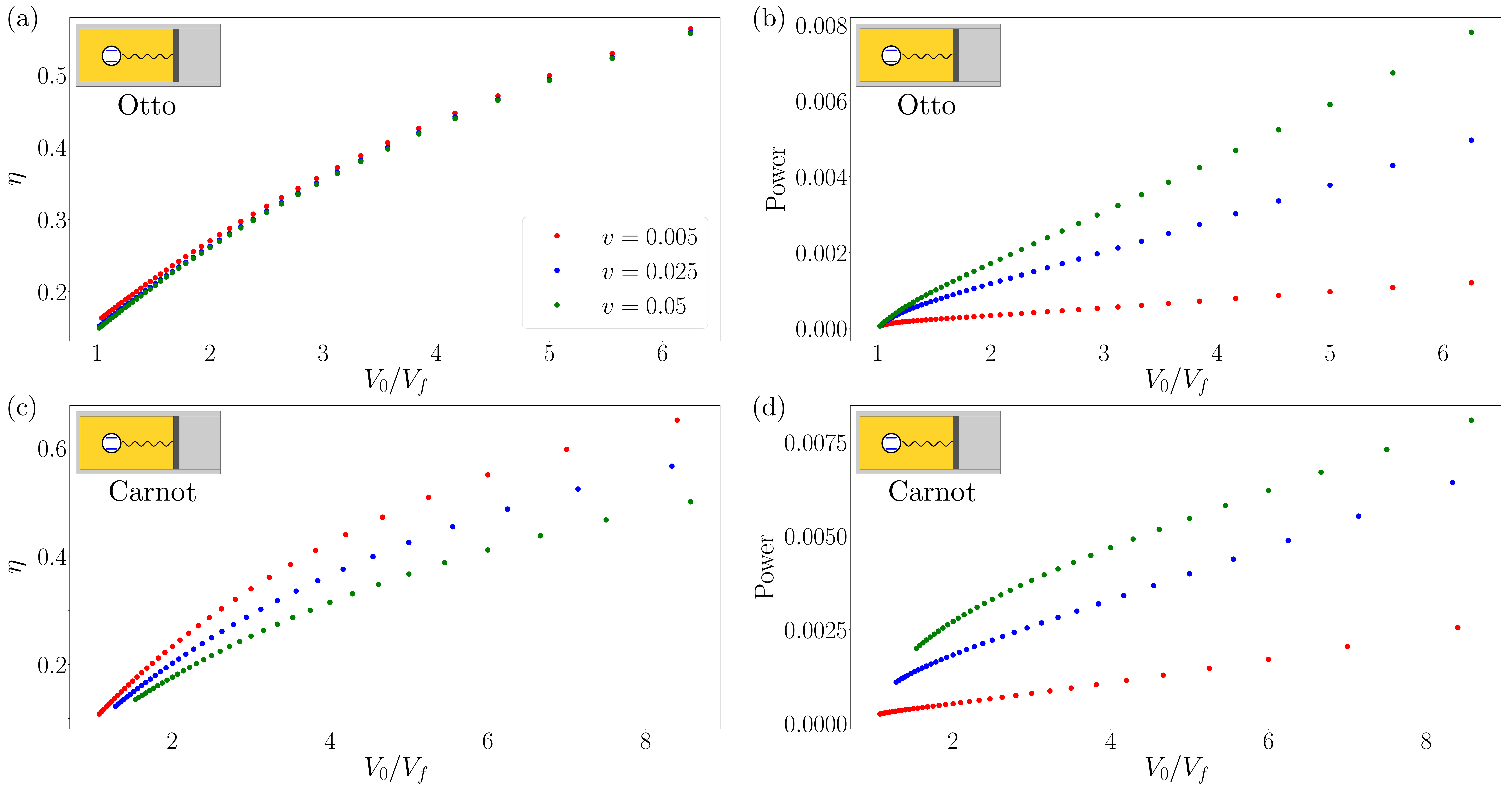}
\vspace{-0.3cm}
\caption{(Color online) Efficiency and power output of Carnot and Otto cycles as a function of the compression ratio $V_0/V_f$ for different cavity wall speeds $|v|$ in the case of a thermal bath coupled with the cavity. The left column shows $\eta$ vs $V_0/V_f$ for the Otto (a) and Carnot (c) engines, while the right column displays the power output curves for the Otto (b) and Carnot (d) cycles. In all cases the parameters are $\ave{n}_{c} = 5$, $\ave{n}_{h} = 20$, $\Gamma = 0.01$, $L_1=12.5$, $\omega_C(0) = \omega_A = 2\pi/L_1$, and $\kappa = 10^{-3}$. Note also that the legend in panel (a) applies to all panels.
} 
\label{fig:effpower}
\end{figure*}

Next we explore the effect of finite-time operations on the performance of the different engines. This effect can be inferred from Figs.~\ref{fig:effpower}(a) and ~\ref{fig:effpower}(c), which show the efficiency of the Otto (a) and Carnot (c) engines as a function of the compression ratio $V_0/V_f$, as measured for different cavity wall speeds $|v|$ in the case of thermal bath coupled to the cavity radiation. This shows that, as expected, a faster moving cavity wall leads to a decreasing global efficiency. However, it is interesting to note that, while for the Otto engine the efficiency drop with $|v|$ is maximal for small compression ratios, for the Carnot engine it is the other way around: the efficiency drop with $|v|$ increases with $V_0/V_f$. The performance of a motor is not only captured by its efficiency, there exist other possible figures of merit, such as, e.g., the engine power output, i.e., the amount of work per unit time it can deliver. Figures ~\ref{fig:effpower}(b) and ~\ref{fig:effpower}(d) show the power output of Carnot and Otto cycles as a function of $V_0/V_f$ for different cavity wall speeds $|v|$ when the thermal bath is coupled with the cavity. Remarkably, while the efficiency decreases with the cavity wall velocity, the power output increases with $|v|$ in both the Otto and Carnot cases, the more the larger $V_0/V_f$ is; see Figs.~\ref{fig:effpower}(b) and ~\ref{fig:effpower}(d). In this way we expect a near-optimal engine configuration for a non-trivial compression ratio, with a high combined efficiency and power output. 

Similar results are obtained in the case of thermal baths coupled to the atom, though as noted previously the efficiency levels reached are comparatively lower (data not shown). Instead, the power output in the bath-atom coupling case is similar in all cases to that observed in the bath-cavity coupling case; see Fig.~\ref{fig:effpower}.b,d.

\subsection{Role of stroke protocols}

In our previous analysis, we have simplified the stroke protocols by assuming that (i) each stroke has the same duration $\tau$, and (ii) the cavity wall expands or compresses linearly in time, with a constant speed $|v|$. To study the engine performance in a more general framework, and in particular the role of different stroke protocols, we now generalize the motion of the cavity wall to allow for acceleration. To do so, we introduce a power-law time dependence for the cavity wall position, 
\be
L(t)=\left(L_0^\gamma + vt \right)^{1/\gamma} ,
\ee
with $v=(L_1^\gamma-L_0^\gamma)/\tau$, such that the initial and final positions are fixed, $L(0)=L_0$ and $L(\tau)=L_1$, for a stroke of duration $\tau$. In this way, for $\gamma=1$ we recover the already-studied linear, constant-speed cavity wall evolution, while values $\gamma\ne 1$ lead to an accelerated cavity wall motion in between the fixed extremal points. This implies in particular that, since the optical mode frequency is inversely proportional to the cavity length, $\omega_C(t) \propto L(t)^{-1}$, we have a dependence $\omega_C(t) \sim t^{-1/\gamma}$, so for $\gamma=-1$ the mode characteristic frequency varies linearly in time, allowing us to make contact with quantum engine models based on ions captured in optical traps with linearly-varying trapping frequency \cite{zheng:pre16}.

Inspired also by these models, we additionally explore the role of time-asymmetric protocols on the efficiency of our atom-doped photon engines. In particular, we will focus our analysis on the Otto cycle, which is somewhat simpler to study. For this type of cycle, we will study the dependence of the efficiency on the relative time duration of the conjugate strokes. As discussed in Sec. \ref{sec:cycles}, the Otto cycle consists of two adiabatic (unitary) compression/expansion strokes and two isochoric heating/cooling strokes. The combined duration of each pair of conjugated strokes (adiabatic compression/expansion or isochoric heating/cooling) is $2\tau$. The idea now is to introduce a time asymmetry in the relative duration of each stroke in the pair, while keeping constant the total duration of the pair of conjugated strokes. In this way, the duration of the first stroke (adiabatic and unitary compression) will now be $\tau_1=\alpha_u 2\tau$, while the duration of the conjugated third stroke (adiabatic expansion) will be $\tau_3=(1-\alpha_u) 2\tau$, with $\alpha_u\in[0,1]$ a parameter measuring the time asymmetry between these pair of conjugated strokes. Similarly, the duration of the second stroke (isochoric heating) will be $\tau_2=\alpha_h 2\tau$, while the duration of the fourth stroke (isochoric cooling) will be $\tau_4=(1-\alpha_h) 2\tau$, with $\alpha_h\in[0,1]$. Note that the total duration of the cycle ($t_\textrm{fin}=4\tau$), as well as the duration of both pairs of conjugated strokes ($2\tau$), remain constant $\forall~\alpha_u,\alpha_h$. In this way a large (small) value of $\alpha_u$ means a slow (fast) adiabatic compression stroke, and a conjugated fast (slow) adiabatic expansion stroke. In a similar manner, a large (small) value of $\alpha_h$ means a slow (fast) isochoric heating stroke, and a conjugated fast (resp. slow) isochoric cooling stroke.

\begin{figure}
\includegraphics[width=\linewidth]{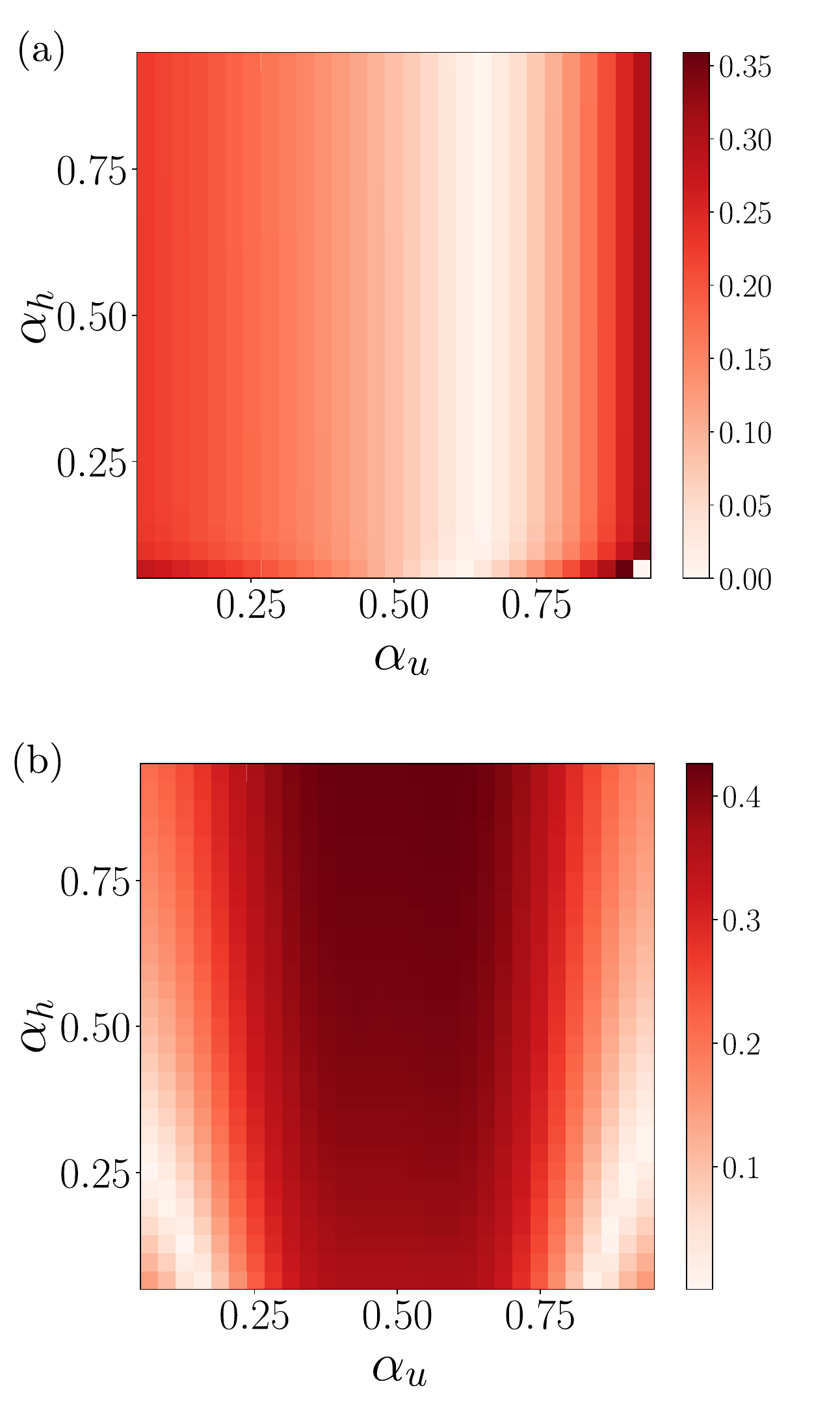}
\caption{(Color online) Efficiency of the Otto cycle measured as a function of parameters $\alpha_u, \alpha_h\in [0,1]$ for acceleration parameter $\gamma=1$ (a) and $\gamma=-1$ (b). Measurements are performed in the quasistatic (small $|v|$) limit and for the case of thermal bath coupled to the cavity mode. The parameters are $\ave{n}_{c} = 5$, $\ave{n}_{h} = 20$, $\Gamma = 0.01$, $L_1=10$, $L_2=5$ (so the compression ratio is $V_0/V_f=2$), $\omega_C(0) = \omega_A = 2\pi/L_1$, and $\kappa = 10^{-3}$. A large $\alpha_u$ ($\alpha_h$) means a slow adiabatic compression (isochoric heating) stroke, and a conjugated fast adiabatic expansion (isochoric cooling) stroke.
} 
\label{fig:finite_time}
\end{figure}

Figure ~\ref{fig:finite_time} shows the efficiency of the Otto cycle as a function of the time asymmetry parameters $\alpha_u, \alpha_h\in [0,1]$ for acceleration parameter $\gamma=1$ [top panel (a)] and $\gamma=-1$ [bottom panel (b)], as obtained in the quasistatic (small $|v|$) limit. For simplicity, we only display the case of the baths coupled to the cavity mode, though the results for bath-atom coupling are qualitatively similar. For the protocol used in previous sections, i.e. the case of uniform (linear in time) motion of the cavity wall [$\gamma=1$, Fig.~\ref{fig:finite_time}(a)], the engine efficiency shows a weak dependence on $\alpha_h$ (except for extremely small values of $\alpha_h$) and a stronger dependence on $\alpha_u$. This implies that the asymmetry in thermalization times during isochoric strokes is less relevant than the asymmetry in the unitary expansion/compression stroke times to optimize efficiency. Moreover, the efficiency dependence on $\alpha_u$ is remarkably asymmetric, showing that the Otto engine becomes more efficient when the compression stroke is performed much faster than the expansion stroke, i.e., $\alpha_u\ll 1$, or vice versa, i.e. $\alpha_u \gg 1$. This observation is mostly independent of the asymmetry in thermalization times during isochoric strokes ($\alpha_h$). Note also that the case discussed in previous sections corresponds to the symmetric choice $\alpha_u=0.5=\alpha_h$.

Figure~\ref{fig:finite_time}(b) shows the efficiency results for $\gamma=-1$, corresponding to a cavity wall motion subject to acceleration. These results are comparable to those in Ref. \cite{zheng:pre16} for an Otto engine model based on an ion captured in an optical trap with a linearly varying trapping frequency. As in the uniform piston motion case, the efficiency for $\gamma=-1$ exhibits a weak dependence on $\alpha_h$ and a steeper dependence on $\alpha_u$, hence implying that the key asymmetry parameter to optimize the engine efficiency is the asymmetry in the unitary expansion/compression stroke times. Interestingly, the efficiency $\eta$ exhibits a bimodal dependency of the efficiency on $\alpha_u$, with two local maxima somewhat below and above $\alpha_u=0.5$, where $\eta$ develops a local minimum. In this way, the optimal efficiency is achieved in a region where the isoparametric expansion and compression strokes are similar in duration, but not equal. We have explored other values of the acceleration parameter $\gamma$ (not shown), concluding that the overall (qualitative) behavior described above is generally dependent on the sign of the parameter $\gamma$, but not on its particular value. In any case, further study is required to fully understand this behavior.

\section{Discussion}
\label{sec:conclusions}

In this paper, we have introduced a model of a microscopic engine that allows us to extract mechanical work from an open quantum system. The model is based on an atom-doped optical quantum cavity that drives a classical piston through radiation pressure. The atom-cavity system is described in terms of the  Jaynes-Cummings model of quantum electrodynamics \cite{jaynes:procieee63,meher:epjp22}, and its dynamics is described by a Lindblad-type master equation \eqref{eq:me} and \eqref{eq:diss} for the atom+cavity interacting system. We demonstrate that the injection of thermal energy into either the cavity or the atom, via an appropriate dissipative coupling to heat baths at different temperatures, can effectively generate work through piston displacements. 

In particular, since the interaction between the system and the moving cavity wall occurs through the radiation pressure generated by the field mode, a first step has been to compute the average expansion work due to this generalized force. We have shown analytically that the expansion work is equivalent (up to first order in the weak atom-radiation coupling limit) to the standard definition of quantum work by Alicki, which is based on the energy change due to time variations in the system Hamiltonian \cite{alicki:jpa79}. This equivalence has been shown to hold in the quasistatic limit, specifically assuming that the state of the system remains thermal throughout the entire process, though our numerical results strongly support that the equivalence extends also to finite-time processes.

Next we have used this setting to build quantum Otto and Carnot engines, analyzing in detail their performance. In particular, we obtain energy-frequency and pressure-volume cycle diagrams, and we compare the energetics, work output, and efficiency of both engines in the quasi-static limit and under finite-time protocols. The Otto cycle consists in an adiabatic compression stroke, followed by an isochoric heating step, an adiabatic expansion stroke, and a final isochoric cooling period. On the other hand, the Carnot engine starts with an adiabatic compression stroke, followed by an isoparametric (constant $\ave{n}$) expansion in contact with a hot bath, an adiabatic expansion step, and another isoparametric compression stroke coupled to a cold reservoir. Initially, the four different strokes for all cases have the same duration $\tau$, and the cavity wall moves at constant speed $|v|$, though we also explored the role of time-asymmetric protocols and accelerated piston motion. 

We found that, when the thermal bath is coupled to the optical mode, the large thermal capacity of the cavity radiation allows us to reach higher energy levels during the heating strokes, in contrast with the lower energy states reached when the thermal bath is coupled to the two-level atom. This results in a much higher work output and efficiency for both types of motors in the bath-cavity coupling cases when compared to the equivalent bath-atom coupling scenarios. In a similar manner, the slower the cavity wall motion, the more time the working medium (atom+radiation) has to drain energy from the hot bath, thus leading to higher energy levels and work outputs. This can also be indirectly corroborated by noticing that the area enclosed by the different pressure-volume cycles decreases as the velocity of the cavity wall increases. A comparison between both types of engines also shows that, for a given cavity wall speed, the Carnot cycle yields a higher work output than the equivalent Otto cycle. To assess the global performance of both types of engines, we also studied both their efficiency and power output as a function of the compression ratio for different cavity wall velocities. Our results show that, while efficiency decreases with an increasing piston speed, the power output increases with $|v|$, suggesting the existence of a near-optimal engine configuration for a nontrivial compression ratio for each piston speed. Finally, the role of time-asymmetric stroke protocols and the accelerated/decelerated motion of the piston have been investigated using efficiency as a figure of merit. We have found that efficiency in this time-asymmetric and accelerated scenario depends mostly on the asymmetry in the unitary expansion/compression stroke times. Moreover, the optimal protocol depends indeed on the acceleration parameter for the cavity wall.

This work has shown how to actually extract useful work from a quantum heat engine to generate net motion. Some questions remain open, such as, for instance, the role of friction on the piston motion, or its actual dynamical coupling with the radiation pressure via classical equations of motion. It would also be interesting to further investigate whether the expansion work and Alicki's definitions of work (and heat) in the quantum realm remain equivalent under more realistic conditions. We plan to address these questions in future work. 

\vspace*{0.5cm}

\begin{acknowledgements}
The research leading to these results has received funding from the fellowship FPU20/02835 and from the Projects of I+D+i Ref. PID2020-113681GB-I00, Ref. PID2021-128970OA-I00, C-EXP-251-UGR23, financed by MICIN/AEI/10.13039/501100011033 and FEDER “A way to make Europe”, and Projects Ref. A-FQM-175-UGR18, Ref. P20\_00173 and Ref. A-FQM-644-UGR20 financed by the Spanish Ministerio de Ciencia, Innovación y Universidades and European Regional Development Fund, Junta de Andalucía-Consejer\'{\i}a de Econom\'{\i}a y Conocimiento 2014-2020. We are also grateful for the computing resources and related technical support provided by PROTEUS, the supercomputing center of the Institute Carlos I for Theoretical and Computational Physics in Granada, Spain. Finally, this work has also been financially supported by the Ministry for Digital Transformation and of Civil Service of the Spanish Government through the QUANTUM ENIA project call - Quantum Spain project, and by the European Union through the Recovery, Transformation and Resilience Plan - NextGenerationEU within the framework of the Digital Spain 2026 Agenda. 
\end{acknowledgements}

\bibliography{./phys.bib}

\end{document}